\documentclass[aps,prb,amsmath,amssymb]{revtex4-2}
\usepackage{bm}
\usepackage{color}
\usepackage{graphicx}
\usepackage{bbold}
\newcommand{\bq}{\begin{equation}}
\newcommand{\ee}{\end{equation}}

\begin{document}
\title{Lattice deformation induced by hot carriers in graphene}

\author{E. G. Mishchenko}
\affiliation{Department of Physics and Astronomy, University of Utah, Salt Lake City, UT 84112, USA}

\begin{abstract}
Hot electrons formed in a graphene crystal by high-intensity short-duration laser pulses  
can exist for a time that is less than an electron-phonon energy relaxation time. During that time, electron-electron collisions cause the electrons to thermalize to a local effective temperature that propagates (diffuses) through graphene. The non-uniform nature of the electron temperature leads to a force acting on the graphene lattice. This force is the result of the electron-lattice interaction that exists even at times that are less than the electron-phonon scattering time. The force causes the lattice to deform. A Boltzmann equation description of the  transient electron-lattice deformation causes by hot electrons in graphene is presented.
\end{abstract}

\maketitle

\section{Introduction}

Physics of optical and transport phenomena related to hot carriers in graphene \cite{MSP} and various other materials is of interest both from a fundamental perspective and numerous potential applications \cite{BHN,KMA,RSM,YBB,AJD}. Such applications include photodetection \cite{KMA,RSM,BLS,TJW,LZC,BL,LSJ}, with photons exciting a photovoltage or photocurrent when photons are absorbed near a boundary between two materials, electro-optical modulators \cite{LYU,PDC,SMC} that modulate electromagnetic waves responsive to electrical signals (voltage) applied to graphene-based devices, non-linear frequency converters \cite{SWR,JHC,ASM}, light-emitting diodes \cite{FCS,BHM,KKC,KGS,KSM,BZM,SGT,LFP} implemented via voltage-biased graphene, and other applications.

One of the ways to create hot electron carriers in graphene is by irradiating a graphene sample with high-intensity short-duration, e.g., $\sim 10^{-14}-10^{-13}$ s, laser pulses. A short high-intensity pulse is mostly absorbed by $\pi$-band conduction electrons of graphene. Electron-electron interaction in graphene is weakly screened while the effective charge $e^2/\hbar v$ is relatively large (of the order of unity), where $v\approx 1.1\cdot 10^6$ m/s is the Dirac velocity of graphene electrons \cite{KUP}. As a result, the electron-electron scattering rate $\tau_{ee}\sim T^{-1}$ is determined by the temperature of the electrons. At temperatures of $T\sim 0.1$ eV, the mean-free time of electron-electron scattering is, therefore, $\tau_{ee}\sim 10^{-13}$ s, which is an order of magnitude less than the typical time of electron-lattice thermalization \cite{TSJ,ZSL,SL,OB}, a phenomenon having a known counterpart in conventional metals \cite{KLT}. Additionally, heat capacity of the lattice is significantly below that of the electrons. As a result, a transient two-temperature state is formed (whose duration is determined by the electron-phonon interaction) in which electrons remain at a temperature $T$ that is substantially higher than the lattice temperature $T_0$, a phenomenon first predicted theoretically to occur in conventional metals in Ref.~\cite{ABE,AKP}.

In addition to electron cooling due to the electron-phonon interaction, two other phenomena take place. First, the heat absorbed by the electrons from the laser pulse propagates via the thermoconductivity mechanism limited by the electron diffusion. Second, non-uniform hot-electron temperature $T({\bf r},t)$ exerts an elastic force on the crystal lattice \cite{FM}. The lattice, therefore, experiences an influence of the hot electron carriers well before a noticeable amount of energy is transferred to the lattice via the electron-phonon interaction. Such non-uniform heating may cause ablation of the crystal lattice with light pulses that are insufficiently energetic to melt the lattice directly.

In this paper, we consider the dynamics of the lattice deformation caused by  hot electrons in graphene heated by short laser pulses. We show that the density of the elastic force exerted by the hot electrons on the lattice is determined by the gradient of the cube of the electron temperature,
\begin{equation}
    \label{force_result}
    {\bf G}({\bf r},t) \sim \frac{\hbar^2 }{v^{2}}\,\nabla T^3 ({\bf r},t),
\end{equation}
and estimate the lattice deformation caused by such force under different regimes of electron transport.

The paper is organized as follows,
 In Section II, we formulate the Boltzmann equation approach to a non-equilibrium state of graphene that includes interaction of electron distribution with lattice deformations. In Section III, we obtain the solution of the Boltzmann equation in the limit where the electron temperature is substantially higher than the lattice temperature. In Section IV, we evaluate lattice deformation that occurs as a result of heating of the electron subssystem in various ranges of distances and times from the laser pulses that are used to heat the system.  
 
\section{Boltzmann equation description of the out-of-equilibrium graphene}

We consider physics that occurs over short times following fast laser heating of electrons of graphene. A monolayer of graphene is assumed to be suspended in air or vacuum. Initially, the graphene sample  is assumed to have a low temperature $T_0$, e.g., room temperature, $T_0 \sim 25\, \text{meV}$. (For conciseness, the system of units is used in which $\hbar=k_B=1$.) A laser pulse of a much higher frequency $\omega \gg T_0$  is incident on the graphene perpendicularly to its plane. The pulse has intensity   of $I({\bf r},t)=c{E^2_0}({\bf r},t)/8\pi$, where ${\bf E}_0({\bf r},t)$ is the (time- and position-varying) amplitude of the electric field, ${\bf E}({\bf r},t)={\bf E}_0({\bf r},t) \cos{(\omega t)}$.

Assuming that $\omega$ is within the range of 2-3 eV, i.e. substantially less than the bandwidth of the $\pi$-band of graphene, the electron spectrum in the Dirac approximation is linear,
\begin{equation}
  \label{spectrum}
\epsilon_{\pm}({\bf p}) =\pm vp,  
\end{equation}
where the upper and lower signs correspond to the upper and lower Dirac cones, respectively.

The electric field of the laser pulse causes vertical transitions from filled states with momentum $p=\omega/2v$ and energy $\epsilon=-\omega/2$ to empty states with the same momentum and energy $\epsilon=\omega/2$. The probability of such transitions can be calculated using the Golden rule for perturbation Hamiltonian, 
\begin{equation}
 V=-\frac{e}{c} {\bf A}\cdot  \hat {\bf v}  =-\frac{e{\bf E}_0}{\omega} \cdot \hat {\bf v} \sin {(\omega t)},
\end{equation}
where $\hat {\bf v}$ is the operator of electron velocity and ${\bf A}$ is the vector potential of the pulse.

The laser pulse drives the electrons to a non-equilibrium state that can be described with the distribution function $f_{\bf p}({\bf r},t) = \{{f^{(+)}_{\bf p}({\bf r},t), f^{(-)}_{\bf p}({\bf r},t)}\}$ that includes the distribution function $f^{(+)}_{\bf p}({\bf r},t)$ for the electrons  of positive energies  and the distribution function $f^{(-)}_{\bf p}({\bf r},t)$ for the electrons  of negative energies.
The two sets of electrons have opposite velocities (for the same momenta ${\bf p}$). The Boltzmann equation for the electrons with positive energies has the form (arguments ${\bf r}$ and $t$ in the distribution function suppressed for brevity),
\begin{eqnarray}
\label{Boltzmann_upper}
\partial_t f^{(+)}_{\bf p} + v {\bf n}\cdot \nabla f^{(+)}_{\bf p} &=&\frac{\pi v^2\sin^2\theta}{2\omega^2} e^2E_0^2 \delta (\omega -2vp) (f_{\bf p}^{(-)} -f_{\bf p}^{(+)})\nonumber\\&& +I_{e-e}[f^{(+)}_{\bf p},f^{(-)}_{\bf p}] + I_{e-ph}[f^{(+)}_{\bf p},f^{(-)}_{\bf p}, N_{\bf k}^{(n)}]+I_{i-e}[f^{(+)}],
\end{eqnarray} 
where $\theta$ is the angle that the momentum ${\bf p}$ makes with th edirection of the electric field ${\bf E}_0$.

The first term  the right-hand side of Eq.~\ref{Boltzmann_upper}) amounts for the (Golden-rule) probability of pulse-driven transitions from the lower Dirac cone to the upper Dirac cone. The  $I_{e-e}$ term represents the electron-electron collision integral and describes both the collisions between two electrons of the upper cone as well as the collisions that involve the electron with momentum ${\bf p}$ in the upper cone and an electron in the lower cone. (The specific form of this collision integral will not be needed.) The $I_{e-ph}$ term stands for the electron-phonon collision integral. It describes the processes in which an electron from the upper Dirac cone and momentum ${\bf p}$ emits a phonon with momentum ${\bf k}$ and transitions to an empty state with momentum ${\bf p}-{\bf k}$ in the upper or lower Dirac cone as well as the inverse processes where the electron from the upper or lower Dirac cone with momentum ${\bf p}-{\bf k}$ absorbs a phonon with momentum ${\bf k}$ and transitions into the state with momentum ${\bf p}$ in the upper cone. The function $N_{\bf k}^{(n)}$ stands for the phonon distribution function of phonons of $n${\it th} branch (counting both acoustic and optical phonon branchess) with momentum ${\bf k}$. The actual form of the eletron-phonon collision integral is preseneted in Appendix \ref{appendix:e-ph}. The last term in Eq.~(\ref{Boltzmann_upper}) describes collisions with impurities and other imperfections.

Similarly, the Boltzmann equation for the electrons in the lower Dirac cone reads,
\begin{eqnarray}
\label{Boltzmann_lower}
\partial_t f^{(-)}_{\bf p} -v {\bf n}\cdot \nabla f^{(-)}_{\bf p} &=&-\frac{\pi v^2\sin^2\theta}{2\omega^2} e^2E_0^2 \delta (\omega -2vp) (f_{\bf p}^{(-)} -f_{\bf p}^{(+)})\nonumber\\&& +I_{e-e}[f^{(-)}_{\bf p},f^{(+)}_{\bf p}] + I_{e-ph}[f^{(-)}_{\bf p},f^{(+)}_{\bf p}, N_{\bf k}^{(n)}]+I_{i-e}[f^{(-)}],
\end{eqnarray} 
Although the Boltzmann equations (\ref{Boltzmann_upper})-(\ref{Boltzmann_lower}) are written for a monochromatic laser pulse, a non-zero linewidth of the pulse can be incorporated by replacing the delta-function $\delta (\omega -2vp)$ with an appropirate Lorentzian function. 


The non-uniform heating of graphene by the laser pulse causes deformation  $u_j ({\bf r},t)$ of the lattice of graphene. Deformation is characterized by the strain tensor $u_{jk} =({\partial}_j u_k +\partial_k u_j)/2$. Strain $u_{jk}$ causes the effective Hamiltonian of the electrons in the $\pi$-band of graphene to change. The strongest contribution comes from a uniform shift of the entire $\pi$-band (relative to the $\sigma$-band) that is proportional to the local contraction/dilatation, $D_a u_{jj}$, where $D_a$ is a constant of the order of several electron-Volts \cite{TA}. A second (weaker) contribution arises from the strain causing changes in the length of the carbon-carbon bonds \cite{TA}. This changes the hopping amplitude of the $\pi$-electrons and, accordingly, the bandwidth of the $\pi$-band. At the low-energy portion of the spectrum, such a change amounts to the modification of the Dirac velocity of electrons. (Additionally, strain $u_{jk}$ changes the size of the first Brillouin zone and location of the Dirac points. These two effects, however, do not change the energy of the electrons and will be ignored.) Accordingly, the local and time-dependent electron spectrum of the deformed graphene becomes, 
\begin{equation}
\label{def_energy}
\epsilon_\pm ({\bf p},{r},t)= \pm vp + D_a u_{jj}({\bf r},t) \pm vp \Lambda({\bf n}, {\bf r},t),~~~ \Lambda ({\bf n}, {\bf r},t)=\lambda_{jk}({\bf n}) u_{jk}({\bf r},t),
\end{equation}

The hexagonal symmetry of graphene implies that at low energy, the modification of the energy spectrum must be isotropic with the symetric tensor of deformation potential constants depending only on the direction ${\bf n}={\bf p}/p$ of the electron momentum:
\begin{equation}
\label{def_pot}
\lambda_{jk}({\bf n}) =\lambda \delta_{jk}+ \lambda' (n_jn_k-\delta_{jk}/2).     
\end{equation}
One can expect that by the order of magneitude the constants $\lambda, \lambda' \sim 1$.

\section{The two-temperature solution of the Boltzmann equation}

The electron-electron collisions conserve the total energy of the electrons and drive the electrons to a local equilibrium at a position and time-dependent temperature $T({\bf r},t)$. The frequency of the electron-electron collisions is determined by the temperature (since no other energy scale is present in the problem),
\begin{equation}
\tau_{ee}^{-1}\sim T.
\end{equation}
At temperatures $T\sim 0.5 - 1 \text{eV}$ (which are still substntially smaller than the bandwidth of the $\pi$-band and where the linear Dirac approximation still applies), the characteristic time of electron-electron collisions is $\tau_{e-e} \sim 10^{-14}\, {s}$. This is the time of establishing the local equilibrium in the electron subsystem.

The electron-phonon collision time, which determines the rate at which electron subsystem loses energy to the lattice, is much longer, $\tau_{e-ph} \sim 10^{-12}\, s$ (see Appendix \ref{appendix:e-ph} for details). The phonons, therefore, remain at the initial (and much lower than $T$) temperature $T_0$ for times that are shorter than $\tau_{e-ph}$.

 The local quasi-equilibrium of the electron subsystem allows ones to write the electron  distribution function as,
\begin{equation}
\label{loc}
f_{\bf p}^{(\pm)}({\bf r},t) = \frac{1}{ \exp{(\pm vp (1+\Lambda ({\bf n},{\bf r},t))/T({\bf r},t))}+1} +\delta f_{\bf p}^{(\pm)}({\bf r},t),
\end{equation}
where the first term is the local Fermi-Dirac distribution function and the second term describes non-isotropic departures from the local equilibrium.

We note that the uniform shift of the $\pi$-band, $D_a u_{jj}$, is compensated by the equal shift of the chemical potential. This follows from the fact that the total number of electrons in the $\pi$-band should not affected by the lattice deformation:
\begin{equation}
\label{conserve}
 \int\frac{Nd^2p}{(2\pi)^2}[f_{\bf p}^{(+)}+f_{\bf p}^{(-)}]=0.
\end{equation}
Here $N$ is the total spin-valley degeneracy of the system ($N=4$ in case of graphene).

Substitution of Eq.~(\ref{loc}) into the
Boltzmann equations (\ref{Boltzmann_lower})-(\ref{Boltzmann_upper}) leads to the following equation, to the linear order in the strain $u_{jk}$, with the upper/lower signs corresponding to the upper and lower cones, respectively:
\begin{eqnarray}
\label{kin}
\partial_t \delta f_{\bf p}^{(\pm)}  \pm v {\bf n}\cdot \nabla \delta f_{\bf p}^{(\pm)}&=&
\mp g(vp)  vp \partial_t \Lambda
\mp g(vp)
\frac{vp}{T}\left(\partial_t T  \pm v 
{\bf n}\nabla T \right) +\frac{\pi v^2\sin^2\theta}{2\omega^2} e^2E_0^2 \delta (\omega -2vp) (f_{\bf p}^{(-)} -f_{\bf p}^{(+)})\nonumber\\&& + I_{e-e} [\delta f_{\bf p}^{(\pm)}] +I_{e-ph}[f^{(\pm)}_0, N_{\bf k}^{(n)}] +I_{i-e}[\delta f^{(\pm)}],
\end{eqnarray}
where we introduced the notation 
\begin{equation}
g(x)= \left[4T \cosh^2{\frac{x}{2T}}\right]^{-1}, ~~~ \int\limits_{0}^{\infty} g(x)dx =\frac{1}{2}.   
\end{equation}

The first term in the right-hand side in Eq.~(\ref{kin}) describes the ``driving force'' with which the time-dependent lattice strain causes the electron distribution function to evolve. The second term in the right-hand side similarly captures the effects of the time- and position-dependent electron temperature. The fourth term $I_{e-e} [\delta f_{\bf p}^{(\pm)}]$ represents the linearized (in $\delta f_{\bf p}^{(\pm)}$) electron-electron collision integral (the zeroth order collision integral vanishes). Finally, the last term describes the rate of change of the electron distribution that occurs as a result of collisions of equilibrium electrons at temperature $T$ and phonons at temperature $T_0$. (The linear correction $I_{e-ph}[\delta f^{(\pm)}_0, N_{\bf k}^{(n)}]$, which is small compared to the electron-electron collision term $I_{e-e} [\delta f_{\bf p}^{(\pm)}]$ is neglected.)

The electron temperature $T$ is assigned a definite meaning by imposing the following condition,
\begin{equation}
\label{norm}
\int\frac{Nd^2p}{(2\pi)^2} vp [\delta f_{\bf p}^{(+)}-\delta f_{\bf p}^{(-)}] =0,
\end{equation}
which attributes the total energy of the electrons in the $\pi$-band to the local equilibrium function in Eq.~(\ref{loc}). The condition (\ref{norm}) leads to a heat transfer equation, which can be obtained by
multiplying the two Boltzmann equations (\ref{kin}) by $\pm vp$, respectively, adding  them together, and integrating over all momenta. This yields,
\begin{equation}
\label{tep}
 \partial_t T^3 + T^3 \partial_t \Lambda +\beta^{-1} \nabla \cdot {\bf q}
=\beta^{-1} Q-\gamma T^3,~~~ \beta =\frac{3N \zeta(3)}{2\pi v^2}
\end{equation}
the energy current density is:
\begin{equation} \label{flow}
{\bf q}=v^2 \int\frac{Nd^2p}{(2\pi)^2}{\bf n} p[\delta f_{\bf p}^{(+)}+\delta f_{\bf p}^{(-)}],
\end{equation}
and the  laser power $Q$ absorbed by the electrons is
\begin{equation}
\label{heat}
Q({\bf r}, t) =\frac{4\pi}{c}\sigma I({\bf r}, t), ~~~\sigma=\frac{Ne^2}{16}.
\end{equation}
Note that $\sigma$ is the usual ``minimal'' conductivity of graphene. More accurately, $\sigma$ is the optical conductivity of graphene, which happens to be independent of frequencies above the electron scattering rate and  below those frequencies at which the curvature of the band spectrum becomes noticeable.

The last term in Eq.~(\ref{heat}) describes the loss of energy to the phonon subsystem for electron temperatures exceeding the optical phonon energy (see Appendix \ref{appendix:e-ph} for details).

To obtain the solution of the linearized Boltzmann equation (\ref{kin}), we use the relaxation time approximation, in which $I_{e-e} [\delta f_{\bf p}^{(\pm)}]+I_{e-i} [\delta f_{\bf p}^{(\pm)}]\approx -\delta f_{\bf p}^{(\pm)}/\tau $, and neglect the slow electron-phonon collision term. This reduces the Boltzmann equation to the following form,
\begin{eqnarray}
\label{kin1}
\partial_t \delta f_{\bf p}^{(\pm)}  \pm v {\bf n}\cdot \nabla \delta f_{\bf p}^{(\pm)}+ \frac{\delta f_{\bf p}^{(\pm)}}{\tau}&=&
X^{(\pm)}_{\bf p},
\end{eqnarray}
where we introduced the notation for the time- and position-dependent function,
\begin{equation}
\label{tion}
X^{(\pm)}_{\bf p}=\pm g(vp)  vp \partial_t \Lambda
\mp 
g(vp)\frac{vp}{T}\left(\partial_t T  \pm v 
{\bf n}\nabla T \right).
\end{equation}

\par The solution of Eq. (\ref{kin1}) can now be written in the integral form
\begin{equation}
\label{solu}
\delta f_{\bf p}^{(\pm)}= \int\limits_{-\infty}^{t} dt'
X^{(\pm)}_{\bf p}\left({\bf r}\mp v {\bf n}(t-t'), t'\right)
\, e^{- (t-t')/\tau},
\end{equation}
Substituting this solution into the equation for the density of heat current ${\bf q}$
given by Eq.~(\ref{flow}) and taking the integral over the absolute value of the momentum $dp$, we arrive at the integral expression for the energy current,
\begin{equation}
\label{flow1}
{\bf q}({\bf r},t)=-\frac{3N\zeta(3)}{2\pi v}\int\limits_0^{2\pi} \frac{d\phi_{\bf n}}{2\pi} {\bf n} \int\limits_{-\infty}^{t}
dt'\, e^{- (t-t')/\tau}
\left(\partial_{t'}
+{\bf v}\nabla'\right)
\Theta({\bf r}', t')\Big|_{{\bf r}'\to {\bf r}-v{\bf n}(t-t')}.
\end{equation}
where we introduced the notation $\Theta ({\bf r}, t)= T^3 ({\bf r}, t)$ for the cube of the electron temperature. Using the Fourier transform 
with respect to space and time variables allows to calculate the time integral in the expression (\ref{flow1}), which simplifies to:
\begin{equation}
\label{flow2}
{\bf q}({\bf k},\omega) = -\beta v \Theta ({\bf k},\omega) \int\limits_0^{2\pi} \frac{d\phi_{\bf n}}{2\pi}  \frac{{\bf n}(\omega-vk_{\bf n})}
{\omega-vk_{\bf n}+i\tau^{-1}}, ~~~ k_{\bf n} = {\bf k} \cdot {\bf n}.
\end{equation}
Substituting this result into the heat transfer equation (\ref{tep}), and neglecting the lattice deformation term (assuming that the strain is small, $\Lambda \ll 1$)
we obtain in the Fourier representation,
\begin{equation}
\label{tepf}
-i\beta \left(\omega + i\gamma  +   \int\limits_0^{2\pi} \frac{d\phi_{\bf n}}{2\pi}  \frac{{\bf n}(\omega-vk_{\bf n})}
{\omega-vk_{\bf n}+i\tau^{-1}} \right) \Theta ({\bf k},\omega) 
=\frac{4\pi}{c}\sigma I ({\bf k},\omega),
\end{equation}
where $ I ({\bf k},\omega)$ is the Fourier transform of the intensity of $I({\bf r}, t)$.
Under the most interesting realistic conditions, the electron relaxation time is shorter than the scales that determine temporal and spatial dispersion in the problem, and $\omega\tau \ll 1$ and $kv\tau\ll 1$, and the remaining integral in Eq. (\ref{tepf}) can be calculated to have the diffusion form, $iDk^2$, where $D=v^2\tau/2$ is the electron diffusion coefficient. This allows one to write for the electron temperature,
\begin{equation}
\label{tepf_solution}
\Theta ({\bf k},\omega) =i\frac{4\pi \sigma}{\beta c} \frac{  I ({\bf k},\omega)}{\omega + iDk^2 +i\gamma }.
\end{equation}
We obtain that the dynamics of the cube of the electron temperature $\Theta({\bf r},t)$ is that of the attenuated diffusion across the graphene crystal. Interestingly, the heat transfer equation turns out to be {\it linear} for the cube of the electron temperature (as long as the scattering time $\tau$ is temperature-independent, which happens when impurity scattering dominates the electron-electron scattering).

For a short and narrow Gaussian laser pulse, 
\begin{equation}
\label{Gaussian_real}
I({\bf r},t) =  \frac{{\cal E}}{\pi a^2} \delta(t)\, e^{-r^2/a^2},  
\end{equation}
of a characteristic radius $a$ and having the total energy ${\cal E}$, we have $I ({\bf k},\omega)={\cal E}e^{-k^2a^2/4}$. Substituting this intensity into Eq.~(\ref{tepf_solution}) and calculating the inverse Fourier transform, we obtain in the real space and time,
\begin{equation}
\label{tepf_solution_real}
T^3 ({\bf r},t) \equiv \Theta ({\bf r},t) =\frac{\sigma {\cal E}}{\beta c(Dt+a^2/4)} e^{-r^2/(4Dt+a^2)-\gamma t}.
\end{equation}
The maximum temperature is reached at $t=0$ at the center of the pulse, $r=0$,
\begin{equation}
\label{temp_max}
T_{max} =\left( \frac{4\sigma {\cal E}}{\beta c a^2} \right)^{1/3}.
\end{equation}
We note that the ratio $4\sigma/c$ for graphene is equal to the fine structure constant, $e^2/\hbar c$.

\section{Dynamics of lattice deformation}

The non-uniform heating of the electron subsystem described by Eqs.~(\ref{tepf_solution}) and (\ref{tepf_solution_real}) causes the lattice of graphene to deform. A two-dimensional crystal with hexagonal symmetry behaves in the long-wavelength limit, where theory of elasticity applies, as an isotropic medium. Correspondingly, the deformation of graphene is described by the equation for the crystal deformation ${\bf u}$,
\begin{equation}
\label{elas}
\ddot {\bf u} -s_t^2 \nabla^2{\bf u} -(s_l^2-s_t^2) \nabla (\nabla\cdot {\bf u} )={\bf G}/\rho,
\end{equation}
where $s_l=19.9\, km/s$ and $s_t=12.9\, km/s$ are the longitudinal and transverse sound velocities,and $\rho =7.6\cdot 10^{-7}\, kg/m^3$ is the mass density of the lattice. The force exerted by electrons on the lattice is denoted with ${\bf G}$. This force is caused by a spatial variation of the electron distribution function,
\begin{equation}
\label{force1}
G_j = \frac{\partial}{\partial x_k} \int\frac{Nd^2 p}{(2\pi)^2} vp
\lambda_{jk}({\bf n}) [ f_{\bf p}^{(+)}({\bf r},t) -f_{\bf p}^{(-)}({\bf r},t)].
\end{equation}
The appearance of the force that depends on the non-equilibrium state of electrons is known as the dynamical theory of relativity, see Ref.~\cite{K,FM}. This force follows from the contribution tot the Lagrangian density  of the system, $-N\sum_{\alpha {\bf p}} vp[f_{\bf p}^{(+)}-f_{\bf p}^{(-)}] \lambda_{jk} u_{jk}$, caused by the change (\ref{def_pot}) in the electron energy (\ref{def_energy}). Note tha, since the number of electrons is conserved, cf.~Eq.~(\ref{conserve}), the term $D_a u_{jj}$ does not contribute to the force.

Substituting the local equilibrium distribution function (\ref{loc}) and the first-order correction (\ref{solu}) into Eq.~(\ref{force1}), we  obtain in the Fourier representation,
\begin{equation}
\label{force2}
G_i ({\bf k},\omega)=-\beta k_k \Theta ({\bf k},\omega) \int\limits_0^{2\pi} \frac{d\phi_{\bf n}}{2\pi}  
\frac{\tau^{-1} \lambda_{ij}({\bf n}) }
{\omega-v{\bf k} \cdot {\bf n}+i\tau^{-1}}.
\end{equation}
The force acting on the lattice is a linear function of the cube of the electron temperature.
In the dispersionless limit of long-wavelength variations of temperature, $\omega \tau \ll 1$ and $kl\ll 1$, the force has the local form,
\begin{equation}
\label{force3}
{\bf G}({\bf r},t) =\beta \lambda \nabla \Theta ({\bf r},t).
\end{equation}

In addition to causing temperature gradient-causing strain, the force (\ref{force1}) leads to two additional effects. More specifically, the equilibrium  contributions to the electron distribution function results in the temperature-dependent renormalization (softening) of the sound velocities,
\begin{equation}
s_l^2 \to \widetilde s_l^2 =s_l^2 - \left(\lambda^2 +\frac{\lambda'^2}{4} \right) \frac{\beta\Theta}{\rho}, ~~~  s_l^2 \to \widetilde s_l^2 =s_l^2 - \frac{\lambda'^2}{8}\frac{\beta\Theta}{\rho}.
\end{equation}
The second effect is the attenuation of sound caused by interaction with the electrons.

According to the elastic equation (\ref{elas}), the force (\ref{force3}) causes the following longitudinal deformation of the graphene crystal, 
\begin{equation}
\label{elaf}
{\bf u} ({\bf k},\omega) =\frac{4\pi\sigma \lambda} {c\rho} \frac{ {\bf k} I ({\bf k},\omega)}
{(\omega+iDk^2 +i\gamma)[ (\omega+i0)^2-s_l^2 k^2]}.
\end{equation}
Taking the inverse Fourier transform of Eq.~(\ref{elaf}), we obtain the dilatation/compression of the graphene crystal as a function of time and distance to the axis of the pulse,
\begin{equation}
\label{def}
\Delta (r,t)\equiv \nabla \cdot {\bf u}({r},t) =\frac{2\sigma \lambda {\cal E}} {c\rho} \int\limits_0^\infty kdk J_0(kr) \left(
\frac{k^2 e^{-(Dk^2+\gamma) t}}{ (Dk^2+\alpha)^2+s_l^2k^2 }  +\text{Re}\frac{k e^{-is_lk t}}{s_lk+iDk^2+i\gamma} \right) \, e^{-k^2a^2/4},
\end{equation}
where $J_0(kr)$ is the Bessel function.

Assuming the energy exchange due to electron-phonon relaxation to be a slow process, we can neglect $\gamma$ in Eq.~(\ref{def}) and obtain two folowing  contributions to the lattice deformation (omitting the subscript   in the notation of the longitudinal sound velocity $s_{l}$),
\begin{eqnarray}
\label{parts}
\Delta (r,t) =-
\frac{2\sigma \lambda {\cal E}} {c\rho s} \int\limits_0^\infty  dk \,  J_0(kr) 
\frac{sk[e^{-Dk^2 t}-\cos{(skt)]}+Dk^2\sin{(skt)}}{D^2k^2 +s^2 }e^{-k^2a^2/4}.
\end{eqnarray}
The most significant deformation occurs in two regions: (i) the region directly illuminated by the laser beam, where $r\lesssim a$ and (ii) the propagating front of the induced sound wave, $r\approx st$. Below the consider these two regions in more detail.

\subsection{The central region, $r\approx 0$.}

Setting $r=0$ makes the Bessel function unity in Eq.~(\ref{parts}). The resulting integral 
consists of three contributions, $\Delta (0,t)=\Delta_1 (t)+\Delta_2 (t)+\Delta_{3}(t)$, which are considered below.

The first contribution
\begin{equation}
\label{first_part}
\Delta_1 (t) =-\frac{2\sigma \lambda {\cal E}} {c\rho s D} \int\limits_0^\infty  dk \,   
\sin{(skt)} e^{-k^2 a^2/4}
\end{equation}
includes the integral that converges over the wave vectors $k$ that are the smaller of $1/st$ and $1/a$. Namely, for $st \ll a$, the sine function can be approximated with its argument, and for $st \gg a$, the exponential function can be approximated with unity. This gives,
\begin{equation}
\label{first_part_eval}
\Delta_1(t) = -\frac{2\sigma \lambda {\cal E}} {c\rho s D} 
\left\{\begin{array}{ll} 2st/a^2, & st\ll a,\\
1/st, & st\gg a.\end{array} \right.
\end{equation}

The integral in the second term 
\begin{equation}
\label{second_part}
 \Delta_2 (t) =\frac{2\sigma \lambda {\cal E}s} {c\rho  D} \int\limits_0^\infty  dk \,
\frac{\sin{(skt)} }{D^2k^2 +s^2 }e^{-k^2 a^2/4}.
\end{equation}
 depends on the relative size of  three parameters, $a$, $st$, and $D/s$. If the size of the laser spot is sufficiently large, $a \gg \text{max} (st, D/s)$, the sine function can be approximated with its argument and the denominator can be approximated simply as $D^2k^2 +s^2 \to s^2$.
If the size of the spot is small, $a \ll \text{max} (st, D/s)$, the value of the integral depends on the relative magnitude of $st$ and $D/s$. If $st \gg D/s$, the convergence of the integral is due to the sine function whereas the exponential function can be approximated with unity and the denominator  can again be approximated with $s^2$. If $st \ll D/s$, the sine function can be approximated with its argument, and the integral is logarithmic:
\begin{equation}
\label{second_part_eval}
\Delta_2(t)  =\frac{2\sigma \lambda {\cal E}s} {c\rho  D} \times \left\{\begin{array}{ll} \displaystyle \frac{2t}{sa^2}, & st, \, D/s \ll a,\\ \displaystyle 
\frac{st}{D^2} \ln{\left(\frac{D}{s\,\text{max} (a, st)}\right)}, & a, \, st \ll D/s, \\ \displaystyle 
\frac{1}{s^3t}, & a, \, D/s \ll st.
\end{array} \right.
\end{equation}

The third contribution includes two parts, $\Delta_{3}=\Delta_{3a} (t) +\Delta_{3b} (t)$. The integral in the ``diffusion'' part,
\begin{equation}
\label{third_part_a}
\Delta_{3a}(t) =-\frac{2\sigma \lambda {\cal E}} {c\rho } \int\limits_0^\infty  dk \,k\,
\frac{e^{-Dk^2 t-k^2a^2/4}} {D^2k^2 +s^2 }.
\end{equation}
converges at $k\sim (Dt+a^2/4 )^{-1/2}\gg D/s$. Provided that $Dt+a^2/4 \gg D^2/s^2$, the integral converges before the denominator changes appreciably so that $D^2k^2 +s^2 \approx s^2$. In the opposite limit of short times and focused pulses, $(Dt+a^2/4 )^{-1/2}\ll D/s$, the integral in Eq.~(\ref{third_part_a}) is logarithmic, $\propto \int dk/k$, with the lower cut-off being $s/D$ and the upper cut-off being $(Dt+a^2/4)^{-1/2}$:
\begin{equation}
 \Delta_{3a}(t) =
- \frac{\sigma \lambda {\cal E}} {c\rho}  \left\{\begin{array}{ll} \displaystyle \frac{1}{s^2(Dt+a^2/4)}, & (Dt+a^2/4 )^{-1/2}\gg D/s,\\ \displaystyle 
\frac{2}{D^2} \ln{\left(\frac{D}{s\sqrt{Dt+a^2/4}}\right)}, & (Dt+a^2/4 )^{-1/2}\ll D/s.
\end{array} \right.    
\end{equation}
 The third contribution is
\begin{equation}
\label{third_part}
\Delta_{3b}(t)=\frac{2\sigma \lambda {\cal E}} {c\rho } \int\limits_0^\infty  dk \,k\,
\frac{\cos{(skt)}}{D^2k^2 +s^2 }e^{-k^2a^2/4}= \frac{D}{s^2} \frac{d\Delta_2(t)}{dt},
\end{equation}
and follows from Eqs.~(\ref{second_part})-(\ref{second_part_eval}).
With the help of the above formulas, one can analyze the deformation caused by the pulse at all times after its impact. 

The maximum deformation occurs at such times that $st \sim D/s$. In this range, the deformation is
\begin{equation}
\label{delta1_max}
 \Delta_{max} \sim \frac{\sigma  \lambda {\cal E}} {c\rho D^2} \text{min} \left(1, \frac{D^2}{s^2a^2}\right).                                                                              
\end{equation}
The most straightforward way to obtain this estimates is to set $st=D/s$ in Eq.~(\ref{parts}) and then evaluate the resulting integral in the two limits:  $a\ll D/s$ and $a\gg D/s$.
Using Eq.~(\ref{delta1_max}) we can obtain he estimate of the deformation of the graphene lattice by the order of magnitude to be (restoring the full units):
\begin{equation}
 \Delta_{max} \sim \frac{(k_B T_{max})^3}{\rho v^2 s^2 \hbar^2}.   
\end{equation}
At temperatures $k_bT \sim 2\, eV$, the deformation is estimated to be about one to several percent.

\subsection{The sound wave front, $st \approx r$.}

The most significant deformation in this range comes from the second term in the parentheses of Eq.~(\ref{def}) where the integral accumulates from large values of $k$, where $s$ can be neglected in comparison to $iDk$ in the denominator. Assuming first that the size of the laser spot is small, $st-r \gg a$, and neglecting the $k^2a^2/4$ term in the exponent, we write,
\begin{equation}
\label{second_part_3}
\Delta(r\simeq st) \approx - \frac{2\sigma \lambda {\cal E}} {c\rho sD} \int\limits_0^\infty dk J_0(kr)\sin{skt}  \approx  \frac{2\sigma \lambda {\cal E}} {c\rho sD} \frac{1}{\sqrt{s^2t^2-r^2}}.  
\end{equation}
This expression diverges at $r\to st$ indicating a sharp onset of deformation when the sound wave reaches distance $r$. The divergence is an artifact of the assumption of an infinitely narrow beam, used in derivation of Eq.~(\ref{second_part_3}). A finite size of the beam $a$ eliminates this divergence, which arises from infinitely large values of $k$. The $k^2a^2/4$ term in the exponent cuts off wave vectors $k$ at a large but finite value $k_{max} \sim a^{-1}$. Using the asymptotic form of the Bessel function, we obtain,
\begin{equation}
\label{second_part_3_corr}
\Delta (r=st) \approx - \frac{\sigma \lambda {\cal E}} {c\rho sD}  \frac{1}{\sqrt{\pi r}}
\int\limits_0^\infty \frac{ dk}{\sqrt{k} } e^{-k^2a^2/4} =- \frac{2\sigma \lambda {\cal E}} {c\rho sD}  \sqrt{\frac{2}{\pi ra}} \sim   \frac{T_{max}^3}{\rho v^2 s D} \frac{a^{3/2}}{r^{1/2}}
\end{equation}
The magnitude of the deformation (\ref{second_part_3_corr}) is about $\sim \sqrt{a/r}$ smaller than the maximum deforamtion occurring near the center of the pulse, Eq.~(\ref{delta1_max}).

\section{Summary and conclusions}

\par A transient state of hot electron carriers formed in a graphene crystal by high-intensity short laser pulses can exist for a duration that is less than a typical type of the electron-phonon relaxation. Highly energetic electrons efficiently exchange energy among themselves and thermalize to a local temperature that is substantially different from the temperature of the lattice. The electron temperature diffuses through the graphene crystal resulting in an homogeneous state of the electrons. Unequal heating of electrons results in a force acting on the graphene lattice that is proportional to the gradient of the third power of the effective electron temperature, ${\bf G}\sim \nabla T^3/v^2 $. The driving force induces both the local deformation of the graphene crystal as well as a propagation of sound waves. The dynamics of lattice deformation is a combination of diffusion and sound propagation. The resulting strain in graphene can be of the order of several percent for electrons whose temperature is of the order of 1-2 eV.

\appendix

\section{Electron-phonon collisions}\label{appendix:e-ph}

The collision integral that determines the rate of change of the electron distribution function $f^{(\alpha)}_p({\bf r},t)$ (where $\alpha =\pm 1$) in Eqs.~(\ref{Boltzmann_upper}) and (\ref{Boltzmann_lower})  occurring as a result of absorption and emission of phonons has the form [\onlinecite{LLX}],
\begin{widetext}
\begin{eqnarray}
\label{coll_e_ph}
I_{e-ph}[f_{\bf p}^{(\alpha)}]=
\sum_{n{\bf p'} \alpha'}  w^{(n,\alpha \alpha' )}_{{\bf p}{\bf p'}}\delta(\alpha' p'- \alpha p
+\omega_{|{\bf p}'-{\bf p}|}^{(n)})\left((1-f_{\bf p}^{(\alpha)} f^{(\alpha')}_{\bf p'}N_{{\bf p}-{\bf p}'}^{(n)} -
f_{\bf p}^{(\alpha)}(1-f^{(\alpha')}_{\bf p'})(1+N_{{\bf p}-{\bf p}'}^{(n)})\right)\nonumber\\
+ \sum_{n{\bf p'} \alpha'}  w^{(n,\alpha \alpha' )}_{{\bf p}{\bf p'}}\delta(\alpha' p'- \alpha p-\omega_{|{\bf p}'-{\bf p}|}^{(n)})\left((1-f_{\bf p}^{(\alpha)})f^{(\alpha')}_{\bf p'}(1+N_{{\bf p}'-{\bf p}}^{(n)}) -
f_{\bf p}^{(\alpha)}(1-f^{(\alpha')}_{\bf p'})N_{{\bf p}'-{\bf p}}^{(n)}\right),
\end{eqnarray}
where $w^{(n,\alpha \alpha' )}_{{\bf p}{\bf p'}}$ is the probability of absorption of a phonon of the n$^{\it th}$ branch with wave vector (momentum) ${\bf k}={\bf p}' -{\bf p}$ by an electron that transitions from a state cone $\alpha$ and momentum ${\bf p}$ to a state in cone  $\alpha'$ and momentum ${\bf p}'$. The inverse process of emission of a photon with momentum ${\bf p} -{\bf p}'$ has the same probability, 
\begin{equation}
w^{(n,\alpha \alpha' )}_{{\bf p}{\bf p'}}=\frac{\pi|\langle \alpha ' {\bf p}'|H_{e-ph}| \alpha {\bf p}\rangle|^2}{\rho \omega_{k}^{(n)}},
\end{equation}
where $\langle \alpha ' {\bf p}'|H_{e-ph}| \alpha {\bf p}\rangle$ is the matrix element of electron-photon interaction, and $|{\bf p}'-{\bf p}|\equiv k $ is the phonon momentum. For acoustic photons,
\begin{equation}
w^{(\alpha \alpha' )}_{{\bf p}{\bf p'}} = \frac{D^2 k }{\rho s},
\end{equation}
where $\rho$ is graphene's mass density and is the speed of sound in graphene.
The strength interaction of electrons with acoustic phonons vanishes in the limit of zero transferred momentum $|{\bf p}'-{\bf p}| \to 0$ in which the deformation becomes uniform. For optical phonons, the probability of phonon absorption/emission remains non-zero even in the long-wavelength limit,
\begin{equation}
\label{wop}
w^{(op,\alpha \alpha' )}_{{\bf p}{\bf p'}} = \frac{\pi D_o^2 }{\rho \omega_o},
\end{equation}
where $D_o$ is the optical phonon deformation potential ($D_o\sim 1 eV/nm$) and $\omega_o$ is the optical phonon frequency ($\omega_o \sim 200\text{meV}$).

At times that are smaller than the time of energy relaxation  between the electron and the phonon (lattice) subsystems, the phonon distribution 
function $N_k^{(n)}$ in Eq.~(\ref{coll_e_ph}) can be assumed to have the equilibrium Bose-Einstein form at the initial (e.g., room) temperature $T_{0}$ of the system,
$$
N_k^{(n)}=\frac{1}{ \exp{(\omega^{(n)}_k/T_{0})}-1}.
$$
This allows one to rewrite for the first line in the collision integral, Eq.~(\ref{coll_e_ph}), 
\begin{eqnarray}
\label{col_first_line}
&&\delta(\alpha' vp'- \alpha vp
+\omega_{|{\bf p}'-{\bf p}|}^{(n)})\left((1-f_{\bf p}^{(\alpha)} f^{(\alpha')}_{\bf p'}N_{{\bf p}-{\bf p}'}^{(n)} -
f_{\bf p}^{(\alpha)}(1-f^{(\alpha')}_{\bf p'})(1+N_{{\bf p}-{\bf p}'}^{(n)})\right)
\nonumber \\ &=&\delta(\alpha' v p'- \alpha v p
+\omega_{|{\bf p}'-{\bf p}|}^{(n)})
\left[f_0( \alpha p)-f_0(\alpha' p')\right]
\left[ N_{|{\bf p}'-{\bf p}|}^{(n)}(T) - N_{|{\bf p}'-{\bf p}|}^{(n)}(T_{0}) \right] 
\end{eqnarray}
Assuming that the electron temperature $T$ is significantly above the lattice (phonon) temperature $T_0$, one can neglect $N_{|{\bf p}'-{\bf p}|}^{(n)}(T_0) \ll  N_{|{\bf p}'-{\bf p}|}^{(n)}(T)$. Assuming further that the electron temperature exceeds relevant phonon frequencies, we can further approximate, $ N_{|{\bf p}'-{\bf p}|}^{(n)}(T) \approx T/\omega_{|{\bf p}'-{\bf p}|}^{(n)}$. This implies that the phonon energy is small compared with the typical electron energy, $pv\sim T$. In this approximation, the the last line in Eq.~(\ref{col_first_line}) becomes,
\begin{equation}
\approx \delta_{\alpha \alpha'}\delta(\alpha v p'- \alpha vp+\omega_{|{\bf p}'-{\bf p}|}^{(n)})
\left[f_0(\alpha p)-f_0(\alpha p')\right] \frac{T}{\omega_{|{\bf p}'-{\bf p}|}^{(n)}}.
\end{equation}

Similarly, for the second line in the  collision integral we obtain,
\begin{eqnarray}
&&\delta(\alpha' v p'- \alpha v p-\omega_{|{\bf p}'-{\bf p}|}^{(n)})\left((1-f_{\bf p}^{(\alpha)})f^{(\alpha')}_{\bf p'}(1+N_{{\bf p}'-{\bf p}}^{(n)}) -
f_{\bf p}^{(\alpha)}(1-f^{(\alpha')}_{\bf p'})N_{{\bf p}'-{\bf p}}^{(n)}\right) \nonumber \\
&&\approx \delta_{\alpha \alpha'} \delta(\alpha v p'- \alpha v p-\omega_{|{\bf p}'-{\bf p}|}^{(n)}) \left[f_0(\alpha p)-f_0(\alpha p')\right] \frac{T}{\omega_{|{\bf p}'-{\bf p}|}^{(n)}}.
\end{eqnarray}
\end{widetext}
Both terms taken together determine the rate of change of the electron's quasi-equilibrium distribution function becomes,
\begin{eqnarray}
\label{coll_e_ph1}
I_{e-ph}[f_{\bf p}^{(\alpha)}]=
T \sum_{n{\bf p'}}  \frac{w^{(n,\alpha \alpha )}_{{\bf p}{\bf p'}} }{\omega_{|{\bf p}'-{\bf p}|}^{(n)}} \left[f_0(\alpha p)-f_0(\alpha p')\right]  \Bigl[ \delta(\alpha v p'- \alpha v p-\omega_{|{\bf p}'-{\bf p}|}^{(n)})+ \delta(\alpha v p'- \alpha  v p +\omega_{|{\bf p}'-{\bf p}|}^{(n)}) \Bigr].
\end{eqnarray}
The net energy flow from the electrons to the phonons is determined by the integral $N\sum_{{\bf p} \alpha} \alpha vp I_{e-ph}[f_{\bf p}] $ taken over all electron states. In evaluating this integral using (\ref{coll_e_ph1}), t is convenient to split the integrand into two equal portions and swap ${\bf p}\leftrightarrow {\bf p}'$ in one of those portions. After simple transformations, one obtains,
\begin{eqnarray}
\label{heat_flow_phonons}
\frac{d{\cal E}_{e\to ph}}{dt} =N\sum_{{\bf p} \alpha} \alpha vp I_{e-ph}[f_{\bf p}^{(\alpha)}] &=&-T N\sum_{n\alpha} \sum_{\bf pp'} w^{(n,\alpha \alpha )}_{{\bf p}{\bf p'}} \omega_{|{\bf p}'-{\bf p}|}^{(n)} \delta({p'}-p) g(vp)\nonumber\\ && = -\frac{NT}{v} \int\limits_0^\infty \frac{p^2dp}{(2\pi)^2} g(vp) W(p), ~~~ W(p) = \sum_{ \alpha}\int\limits_{0}^{2\pi} \frac{d\phi_{\bf n}}{2\pi} \int\limits_{0}^{2\pi} \frac{d\phi_{{\bf n}'}}{2\pi}w^{(n,\alpha \alpha )}_{p{\bf n},p{\bf n}'} \omega_{p|{\bf n}'-{\bf n}|}^{(n)}. 
\end{eqnarray}
The temperature dependence of $d{\cal E}_{e\to ph}/{dt}$ on electron temperature $T$ can be obtained by counting the total power of the momentum $p$ in the integrand of Eq.~(\ref{heat_flow_phonons}), since the function $g(vp)$ ``forces'' the electron momentum to be $p\sim T/v$. (The function $g(vp)$ itself brings in a power of $T^{-1}$.)

For acoustic phonons, $W(p) \propto p^2 \propto T^2$, and
\begin{equation}
    \frac{d{\cal E}_{e\to ph}}{dt} \propto - T^5.
\end{equation}
Acoustic phonons dominate the electron-phonon energy exchange at  temperatures $T\lesssim \omega_o$ that are less than the optical phonon frequencies. 

At electron temperatures above the optical phonon frequencies, $T\gtrsim \omega_o$, the main role is played by the optical phonons, for which $W(p)$ is independent of temperature. According to Eq.~(\ref{wop}), $W(p)\approx \pi D_o^2/\rho$, so that the integral in Eq.~(\ref{heat_flow_phonons}) yields,
\begin{equation}
    \frac{d{\cal E}_{e\to ph}}{dt} =-\beta \gamma T^3, ~~~\gamma =\frac{\pi^2 D_o^2}{36\rho v^2 \zeta(3)}.
\end{equation}
By the order of magnitude, in graphene, $\rho \approx 7.\cdot 10^{-7} \, kg/m^2$, $v\approx 1\cdot 10^6~m/s$, and $\gamma \sim 10^{12}-10^{13} \text{Hz}$.


\end{document}